# The Derivation of Failure Event Correlation Based on Shadowing Cross-Correlation

Milad Ganjalizadeh, Piergiuseppe Di Marco

milad.ganjalizadeh@ericsson.com

piergiuseppe.di.marco@ericsson.com

*Abstract*— **In this document we derive the mapping between the failure event correlation and shadowing cross-correlation in dual connectivity architectures. In this case, we assume that a single UE is connected to two gNBs (next generation NodeB).**

We define the radio access network (RAN) reliability, $R_{RAN^{(i)}}$, as the probability of success for sending a packet from the UE to the $i$-th gNB (see [1])

$$R_{RAN^{(i)}} = P(D^{(i)} \leq D_{max})P(P_r^{(i)} \geq P_{th}), \qquad (1)$$

where $P(D^{(i)} \leq D_{max})$ is the probability that the observed delay for a packet is lower than the specified requirement for RAN and $P(P_r^{(i)} \geq P_{th})$, is the probability that the received power from the $i$-th path is higher than a minimum threshold, $P_{th}$. Similarly, $\varepsilon_{RAN^{(i)}}$ is defined as the probability of failure on the $i$th path on the RAN part.

In the following, we derive the failure event correlation, assuming that shadowing correlation is the only reason for having correlated events and the delay requirements in (1) are met on both wireless links.

On a particular wireless link, the measured received power can be modelled as

$$P_r^{(i)} = P_t^{(i)} - P_l^{(i)} - X_{dB}^{(i)}, \qquad (2)$$

where $P_r^{(i)}$, $P_t^{(i)}$ and $P_l^{(i)}$ are the received power, transmit power and path loss (all in dB) on the $i$-th path, respectively. Besides, $X_{dB,i}$ is a Gauss-distributed random variable with mean 0 and variance $\sigma_{dB,i}^2$ representing the variations in received power on the $i$-th RAN due to shadow fading [2]. Therefore, the RAN reliability, in (1), is given by

$$R_{RAN^{(i)}} = P\big(P_{th}^{(i)} < P_t^{(i)} - P_l^{(i)} - X_{dB,i}\big) = P\left(\frac{X_{dB,i}}{\sigma_{dB,i}} < \frac{P_t^{(i)} - P_l^{(i)} - P_{th}^{(i)}}{\sigma_{dB,i}}\right). \qquad (3)$$

Let us define the mean normalized received power above threshold for the $i$-th path in RAN, $\beta_i$, as

$$\beta_i = \frac{P_t^{(i)} - P_l^{(i)} - P_{th}^{(i)}}{\sigma_{dB,i}}, \qquad (4)$$

therefore, using (3) the RAN reliability becomes

$$R_{RAN^{(i)}} = 1 - Q(\beta_i), \qquad (5)$$

where $Q(\beta_i)$ is the Q-function of the mean normalized received power above threshold.

The failure event correlation among two different wireless links, $\rho$, can be derived using bivariate Gaussian distribution as

$$\begin{aligned}
\rho &= \frac{P(X_{dB,1} > \beta_1, X_{dB,2} > \beta_2) - \varepsilon_{RAN^{(1)}}\varepsilon_{RAN^{(2)}}}{\sigma_{I_{RAN_f^{(1)}}}\sigma_{I_{RAN_f^{(2)}}}} \\
&= \frac{\int_{\beta_2}^{\infty}\int_{\beta_1}^{\infty}\frac{1}{2\pi\sqrt{1-\rho_h^2}}e^{\frac{-(X_{dB,1}^2 + X_{dB,2}^2 - 2\rho_h X_{dB,1} X_{dB,2})}{2(1-\rho_h^2)}}dX_{dB,1}\,dX_{dB,2} - Q(\beta_1)Q(\beta_2)}{\sigma_{I_{RAN_f^{(1)}}}\sigma_{I_{RAN_f^{(2)}}}} \\
&= \frac{\frac{1}{\sqrt{2\pi}}\int_{\beta_2}^{\infty}Q\left(\frac{\beta_1 - \rho_h X_{dB,2}}{\sqrt{1-\rho_h^2}}\right)e^{\frac{-X_{dB,2}^2}{2}}dX_{dB,2} - Q(\beta_1)Q(\beta_2)}{\sigma_{I_{RAN_f^{(1)}}}\sigma_{I_{RAN_f^{(2)}}}},
\end{aligned} \qquad (6)$$

where $\rho_h$ represents the shadowing cross-correlation between the two RAN links.

In Table I, we report some examples of the mapping between the two (i.e., shadowing correlation and failure event correlation) when $\varepsilon_{RAN} = 10^{-4}$.

Table I: Examples of the mapping between shadowing correlation, $\rho_h$, and event correlation, $\rho$, for $\varepsilon_{RAN} = 10^{-4}$.

| $\rho_h$ | 0.05 | 0.1 | 0.2 | 0.3 | 0.4 | 0.5 | 0.7 | 1 |
|---|---|---|---|---|---|---|---|---|
| $\rho$ | 0.0001 | 0.0003 | 0.0013 | 0.004 | 0.0101 | 0.0232 | 0.1 | 1 |